\documentclass[11pt,twoside,letterpaper]{article} %% The same for {book}
\usepackage{times,fancyhdr}
\usepackage[dvips]{graphicx}

\usepackage{epsfig}
\usepackage{amssymb}
\usepackage{amsmath}
\usepackage{amsfonts}
\usepackage{amsthm,amscd}
\usepackage{amsbsy}
\usepackage{latexsym}
\usepackage{bm} 
\usepackage{bbm}
\usepackage{url} 			%% Nicely format and linebreak URLs in the bibliography (and elsewhere).
\usepackage{layout}
\usepackage{pslatex}
\usepackage{cite}
\usepackage{fleqn} 		% displayed formulas flush left (default is centered).
\usepackage{makeidx}
\makeindex 						% Creates index at the end of the book (with makeidx).
\usepackage{layout} 	% To see the current values of these dimensions, use the layout package, 
\usepackage{epstopdf}	% Automatically converts EPS files to encapsulated PDF files (using ghostscript).
\usepackage{color}
\usepackage{hyperref}
\usepackage[latin1]{inputenc}
\usepackage[T1]{fontenc}
\usepackage{poligraf} %% Color separation
\usepackage[letter,cam,center]{crop} %% Printing crop-marks
\usepackage{type1cm} 	%% Use scalable, PostScript Type 1 versions of the Computer Modern fonts.
\usepackage{courier}	%% Replace the standard Computer Modern Typewriter font LaTeX uses
\usepackage{lscape} 	% for landscape section
\usepackage{slashed}
\def\bea{\begin{eqnarray}}
\def\eea{\end{eqnarray}}


\makeatletter 
\def\cleardoublepage{\clearpage\if@twoside \ifodd\c@page\else% 
    \hbox{}% 
    \thispagestyle{empty}%
    \newpage% 
    \if@twocolumn\hbox{}\newpage\fi\fi\fi} 
\makeatother

\def\figurename{Figure}
\makeatletter
\renewcommand{\fnum@figure}[1]{\figurename~\thefigure.}
\makeatother

\def\tablename{Table}
\makeatletter
\renewcommand{\fnum@table}[1]{\tablename~\thetable.}
\makeatother

\begin{document}
\title{
{\begin{flushleft}
\vskip 0.45in
{\normalsize\bfseries\textit{Chapter~1}}
\end{flushleft}
\vskip 0.45in
\bfseries\scshape Nuclear Data for Astrophysical Modeling}}
\author{\bfseries\itshape Boris Pritychenko$^1$\thanks{E-mail address: pritychenko@bnl.gov} \\
\\
$^1$National Nuclear Data Center, Brookhaven National Laboratory, Upton, NY 11973, USA
 \\
}
\date{}
\maketitle
\thispagestyle{empty}
\setcounter{page}{1}
% ------- [First Page Running Head] - place it immediately after title! ------
\thispagestyle{fancy}
\fancyhead{}
\fancyhead[L]{In: Book Title \\ 
Editor: Editor Name, pp. {\thepage-\pageref{lastpage-01}}} % needs \label{lastpage-01} on the last page.
\fancyhead[R]{ISBN 0000000000  \\
\copyright~2007 Nova Science Publishers, Inc.}
\fancyfoot{}
\renewcommand{\headrulewidth}{0pt}
%------------------------------------------------------------------------------

%\vspace{2in}
\begin{abstract}

Nuclear physics has been playing an important role in modern astrophysics and cosmology. Since the early 1950's it has been successfully applied for 
the interpretation and prediction of astrophysical phenomena.    Nuclear physics models helped to explain the observed elemental 
and isotopic abundances and star evolution and provided valuable insights on the Big Bang theory.       
Today, the variety of elements observed in stellar surfaces, solar 
system and cosmic rays, and isotope abundances are calculated and compared with the observed values.  Consequently, the overall 
success of the modeling critically depends on the quality of underlying nuclear data that helps to bring physics of macro and micro scales together.  
\newline To broaden the scope of traditional nuclear astrophysics activities and produce additional complementary information, 
I will investigate applicability of the U.S. Nuclear Data Program (USNDP) databases for astrophysical applications.   
EXFOR (Experimental Nuclear Reaction Data) and ENDF (Evaluated Nuclear Data File) libraries have 
 large astrophysics potential; the former library contains experimental data sets while the latter library includes evaluated neutron cross sections. 
ENSDF (Evaluated Nuclear Structure Data File) database is a primary depository of nuclear structure and decay rates information. The decay rates are 
essential in stellar nucleosynthesis calculations, and these rates are evaluated using nuclear structure codes. The structure evaluation codes are 
pure mathematical procedures that can be applied to diverse data samples.  
\newline As an example, I will consider Hubble constant measurements. An extraordinary number of Hubble constant 
measurements challenges physicists with selection of the best numerical value. 
Data evaluation codes and procedures help to resolve this issue. The present work produces 
the most probable or recommended Hubble constant value of 66.2(77) (km/sec)/Mpc. This recommended value is based on the last 
25 years of experimental research and includes contributions from different types of measurements. 
The current result implies (14.78$\pm$1.72) $\times$ 10$^{9}$ years as a rough estimate for the age of the Universe.
\newline A brief review of astrophysical nuclear data needs has been presented. Several opportunities and the corresponding computer tools 
have been identified. Further work will include extensive analysis of nuclear databases and computer procedures for astrophysical calculations.

\end{abstract}

%\noindent \textbf{PACS} 05.45-a, 52.35.Mw, 96.50.Fm.
%\vspace{.08in} \noindent \textbf{Keywords:} Energy of vacuum state.

%% Other situations:
%\noindent \textbf{Key Words}: Stellar nucleosynthesis, nuclear data, cosmological parameters
%\vspace{.08in} \noindent {\textbf AMS Subject Classification:} 53D, 37C, 65P.

\section{Introduction and Overview} 

In the last century, astronomy, astrophysics, and cosmology have evolved from observational and theoretical pursuits into more 
experimental science, with many stellar and planetary processes recreated in physics laboratories and extensively studied. 
Many astrophysical phenomena have been explained using the present understanding of nuclear physics processes, and the whole concept of 
stellar nucleosynthesis has been introduced. Further developments helped to identify the Big Bang, stellar, and explosive nucleosynthesis processes 
that are responsible for the currently-observed variety of elements and isotopes \cite{Burbidge57,Cameron57}. Today, nuclear physics 
is successfully applied to explain the variety of elements and isotope abundances observed in stellar surfaces, the solar system,
and cosmic rays via network calculations and comparison with observed values. 

A comprehensive analysis of stellar energy production, metallicity, and isotope abundances indicates a crucial role of proton, neutron, and 
light ion-induced nuclear reactions and $\alpha$-, $\beta$-decay rates. These subatomic processes govern the observables and predict the star 
life cycle. Calculations of the transition rates between isotopes in a network strongly rely on theoretical and experimental cross section 
and decay rate values at stellar temperatures. Consequently, the general availability of nuclear data is of paramount importance in stellar 
nucleosynthesis research. Application of nuclear data codes and evaluation procedures to Hubble constant measurements shows a high potential 
of nuclear data analysis and mining techniques in astrophysics and cosmology.

This chapter will provide a review of theoretical and experimental nuclear reaction and decay data for stellar and 
explosive nucleosynthesis, as well as modern computation tools and methods. Examples of evaluated and compiled nuclear physics 
data will be given. Applications of nuclear databases and data evaluation methods for nucleosynthesis calculations  and Hubble constant measurements evaluation, respectively, will be discussed.

% ------------ [Running Heads - for odd and even pages] - please insert it only on page 2!
\pagestyle{fancy}
\fancyhead{}
\fancyhead[EC]{B. Pritychenko}
\fancyhead[EL,OR]{\thepage}
\fancyhead[OC]{Nuclear Data for Astrophysics Modeling}
\fancyfoot{}
\renewcommand\headrulewidth{0.5pt} 
%------------------------------------------------------------------------------

\section{Nucleosynthesis Models}

Nucleosynthesis is an important physical phenomenon that is responsible for chemical elements and isotope abundances. 
It started in the early Universe and  proceeds in the stars. The Big Bang nucleosynthesis is responsible for a relatively high abundance 
of the lightest primordial elements in the Universe from $^{1}$H to $^{7}$Li, and it precedes star formation and stellar nucleosynthesis. 
The general consistency between theoretically predicted and observed  abundances of the lightest elements serves as a strong evidence for the 
Big Bang theory \cite{Kolb88}.
\begin{figure}[t]
 \centering
 \includegraphics[width=0.7\textwidth]{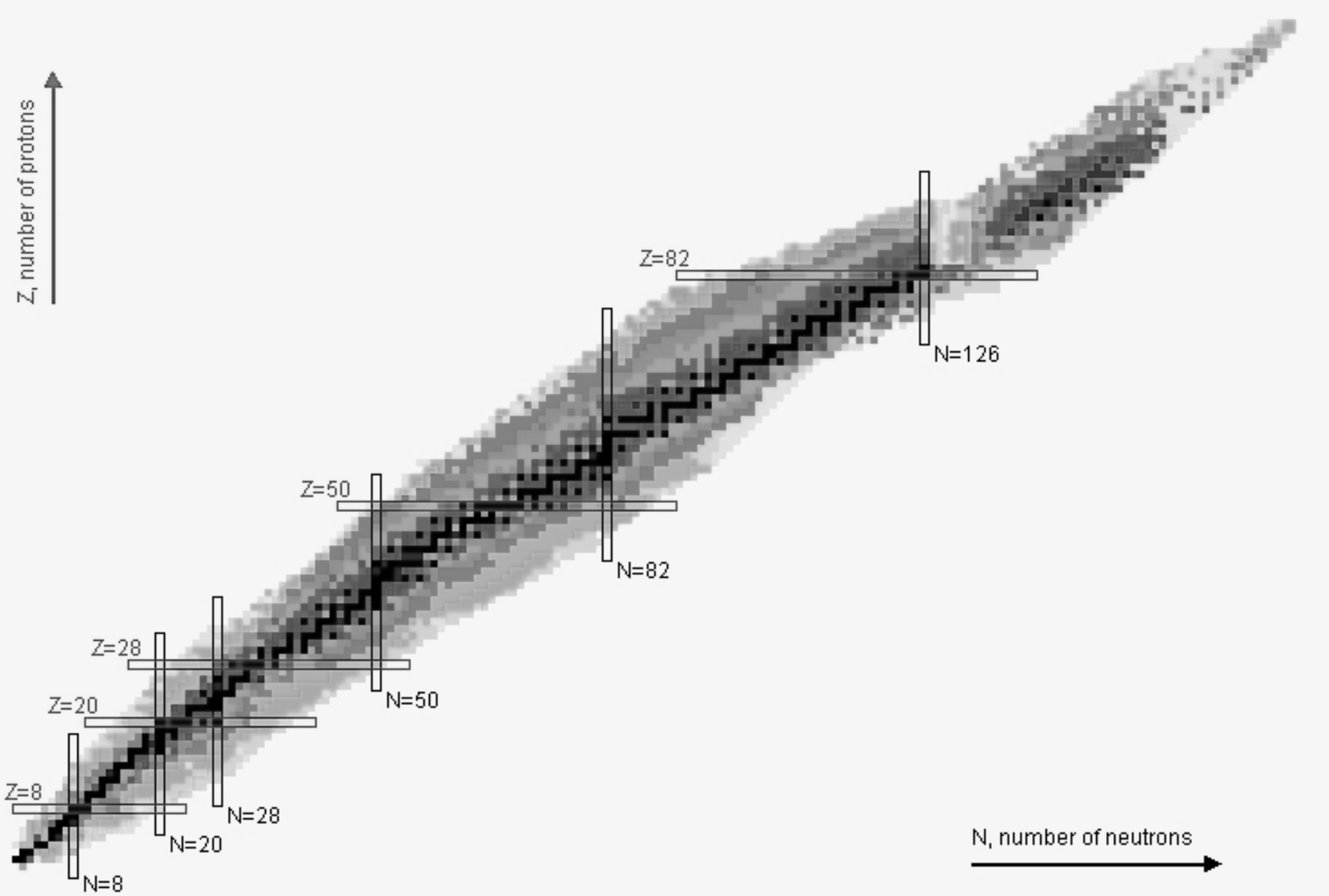}
\caption{The chart of nuclides. Stable and long-lived ($>$10$^{15}$ s) nuclides are shown in black. Courtesy of NuDat Web application 
({\it http://www.nndc.bnl.gov/nudat}).}
\label{fig:chart}
\end{figure}

\begin{figure}[t]
 \centering
 \includegraphics[width=0.7\textwidth]{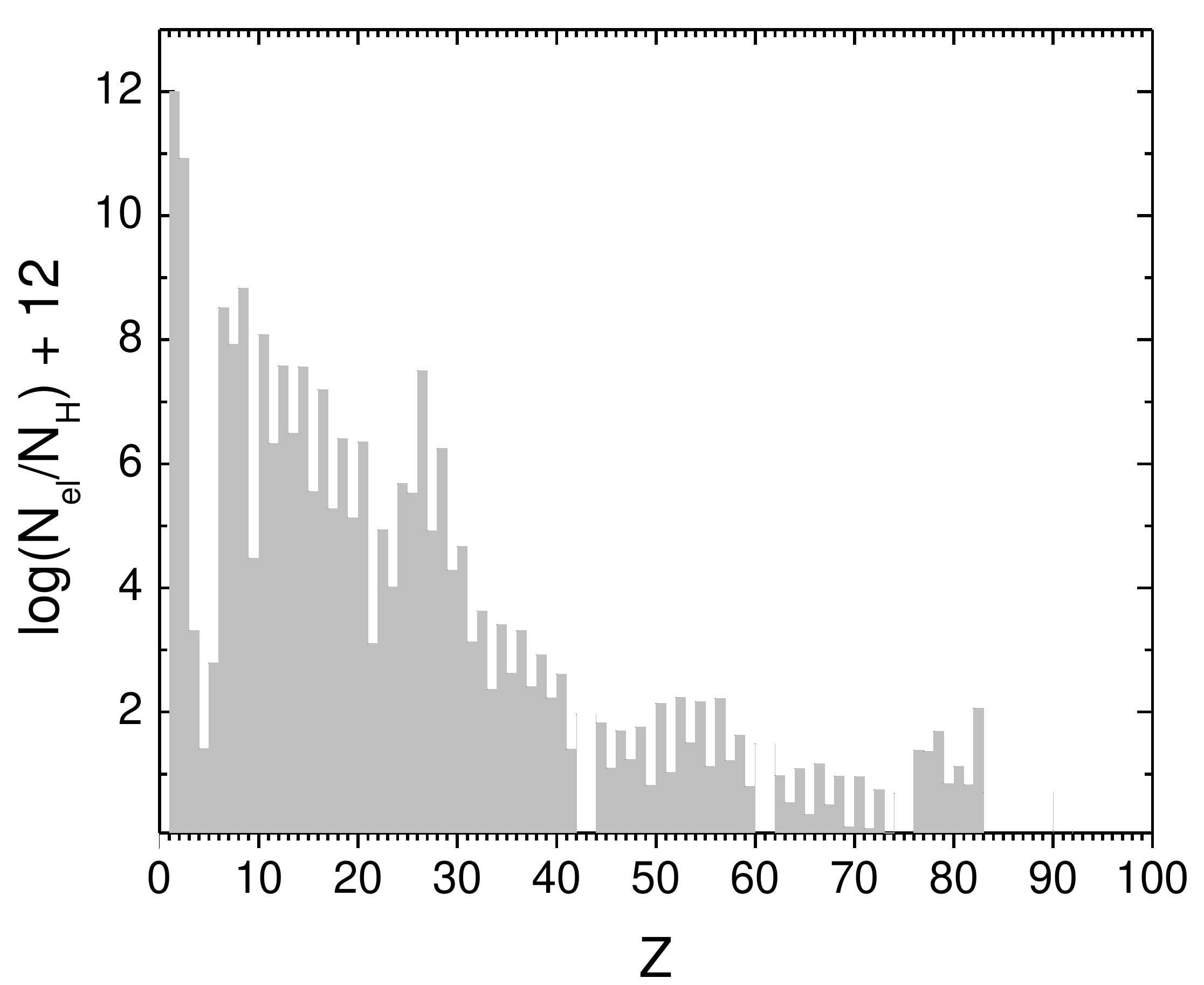}
\caption{Solar system elemental abundances; data are taken from \cite{Grevesse98}.}
\label{fig:solar}
\end{figure}

Presently-known varieties of nuclei and element abundances are shown in Fig. \ref{fig:chart} and Fig. \ref{fig:solar}. 
These Figures indicate a large variety of isotopes in Nature \cite{Anders89} and the strong need for additional nucleosynthesis mechanisms 
beyond the Big Bang theory. These additional mechanisms have been pioneered by Eddington \cite{Eddington20} via introduction of a revolutionary 
concept of element production in the stars. Present nucleosynthesis models explain medium and heavy element abundances using stellar 
nucleosynthesis, consisting of burning (explosive) stages of stellar evolution, photo disintegration, and neutron and proton capture processes. 
These model predictions can be verified through  star surface metallicity studies and comparison of calculated isotopic/elemental abundances with 
 observed values, as shown in Fig. \ref{fig:solar}.

A detailed analysis of elemental abundances and radioactive decay rates shows several mechanisms for production of currently-observed 
chemical elements and isotopes. The slow-neutron capture ({\it s}-process) is responsible for creation of about 50 $\%$ of the elements beyond iron.  
In this region, neutron capture  becomes dominant because of the increasing Coulomb barrier and decreasing binding energies. 
The {\it s}-process takes place in the Red Giants and AGB stars, where neutron temperature ($kT$) varies from  8 to 90 keV. 
Further analysis of stable and long-lived nuclei indicates the large number of isotopes that lie outside of the {\it s}-process path peaks near A=138 and 208. 
In addition, the large gap between {\it s}-process nucleus $^{209}$Bi and $^{232}$Th, $^{235,238}$U effectively terminates the {\it s}-process at $^{210}$Po. 
Production of the actinide neutron-rich nuclei cannot be explained by {\it s}-process nucleosynthesis and requires introduction of a rapid neutron capture 
or {\it r}-process. The Fig. \ref{fig:chart} data also indicate between 29 and 35 proton-rich nuclei that cannot be produced in the {\it s}- or {\it r}-processes. 
A significant fraction of these nuclei originate from the $\gamma$-process \cite{Boyd08,Woosley78}.

Nowadays, there are many well-established theoretical models of stellar nucleosynthesis \cite{Boyd09,Burbidge57}; however, they still cannot reproduce 
the observed abundances due to many nuclear physics parameters that are not-well defined  and stellar properties  uncertainties. In this chapter, I will concentrate on the better known {\it s}-process nucleosynthesis. 
The flow of nuclear physics processes in the network calculations is defined 
by nuclear masses, reaction, and decay rates, and strongly correlated with stellar temperature and density. These calculations depend 
heavily on our understanding of temperature, density and chemical composition of stars, and the availability of high quality nuclear data.

\section{Nucleosynthesis Data}

There are many sources of stellar nucleosynthesis data that include KADoNiS, NACRE and REACLIB dedicated nuclear astrophysics libraries 
\cite{Angulo99,Cyburt10,Dillmann06}, and the U.S. Nuclear Data Program  databases.

\subsection{Astrophysical Libraries}

The KADoNiS, NACRE and REACLIB dedicated nuclear astrophysics libraries are optimized for nuclear astrophysics applications  
and contain pre-selected data that are often limited to their original scope. In many cases, these data sources reflect 
the present state of nuclear physics, when experimental data are not always available or limited to a single measurement. 
Such limitation highlights the importance of theoretical calculations that are strongly dependent on nuclear models. In addition, 
nucleosynthesis processes are strongly affected by  astrophysical site conditions. To enlarge the 
scope of  traditional nuclear astrophysics calculations, I will investigate applicability of nuclear physics databases for 
stellar nucleosynthesis data mining. These databases were developed for nuclear science, energy production, and national security 
applications and will provide complementary and model-independent results.

\subsection{Nuclear Reaction Data}

The experimental and evaluated nuclear reaction data are archived in EXFOR  and ENDF  libraries, respectively.

\subsubsection{EXFOR}
The EXFOR database \cite{Holden05,Otuka14} contains an extensive compilation of experimental nuclear reaction data. 
The EXFOR library was started in 1967 at a meeting of the four major nuclear data centers: National Nuclear Data Center, 
Brookhaven National Laboratory, USA; NEA Databank, Paris, France; Nuclear Data Section, IAEA, Vienna, Austria; and Nuclear Power Engineering Institute, 
Obninsk, Russian Federation \cite{Holden05}. This project was based on previously existing efforts, such as 
the Brookhaven Sigma Center Information Storage and Retrieval System (SCISRS). Currently, EXFOR is an international collaboration under the auspices of the International 
Atomic Energy Agency  (IAEA), and scientists from the USA, France, Austria, Russian Federation, China, Hungary, India, Japan, Korea, and Ukraine contribute to it.

\begin{figure}[t]
 \centering
 \includegraphics[width=0.7\textwidth]{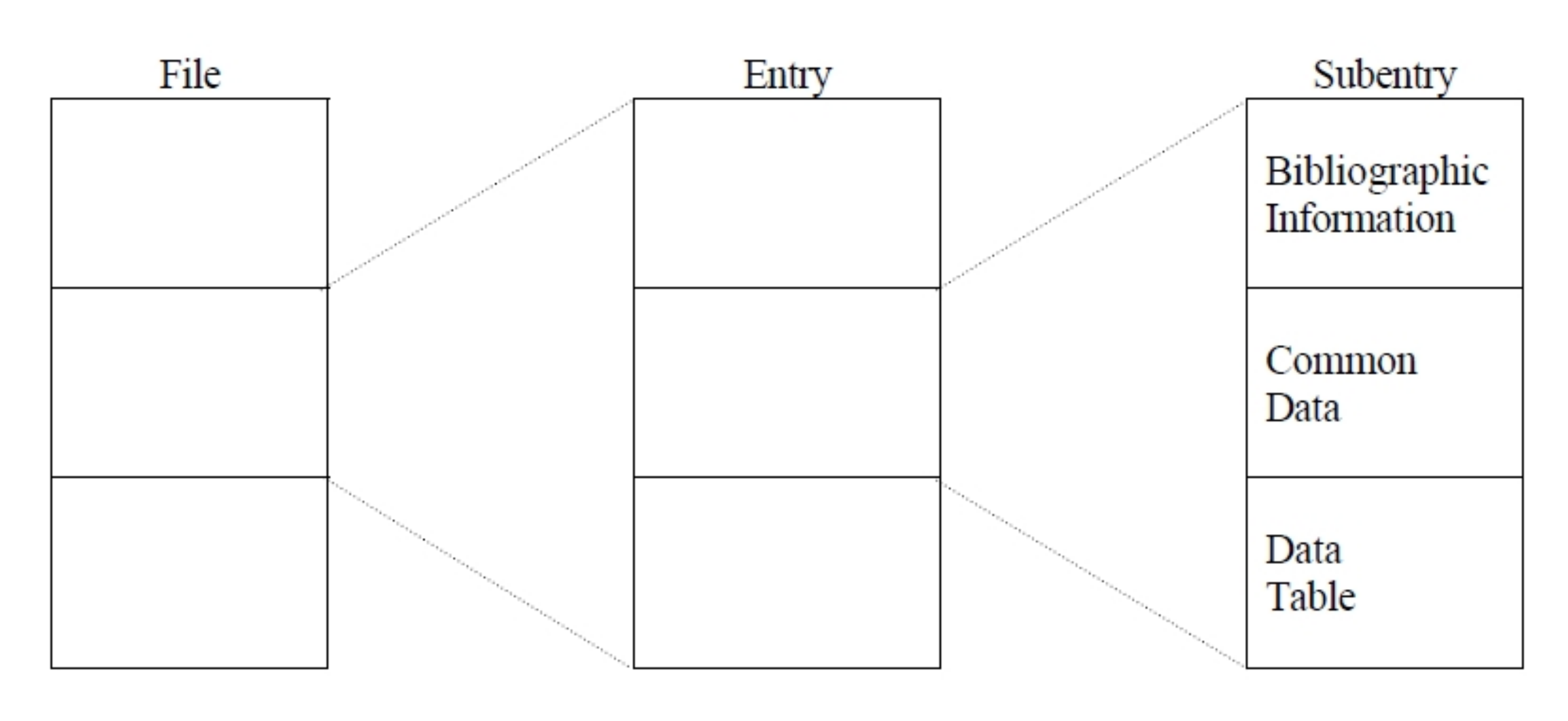}
\caption{Structure of an EXFOR compilation file.}
\label{fig:x4entry}
\end{figure}

EXFOR or EXchange FORmat file was established for  information interchange between the four major data centers,   
and its file structure is shown in Fig. \ref{fig:x4entry}. The library consists of data entries that contain complete records of individual experiments.   
Each experiment may include multiple nuclear reaction data sets (subentries) and several research papers because EXFOR compilers 
group multiple publications from a particular experiment into a single entry. Another important feature of the 
library is that all experimental data are compiled as published, only obvious errors are corrected.

\begin{figure}[]
 \centering
 \includegraphics[width=0.7\textwidth]{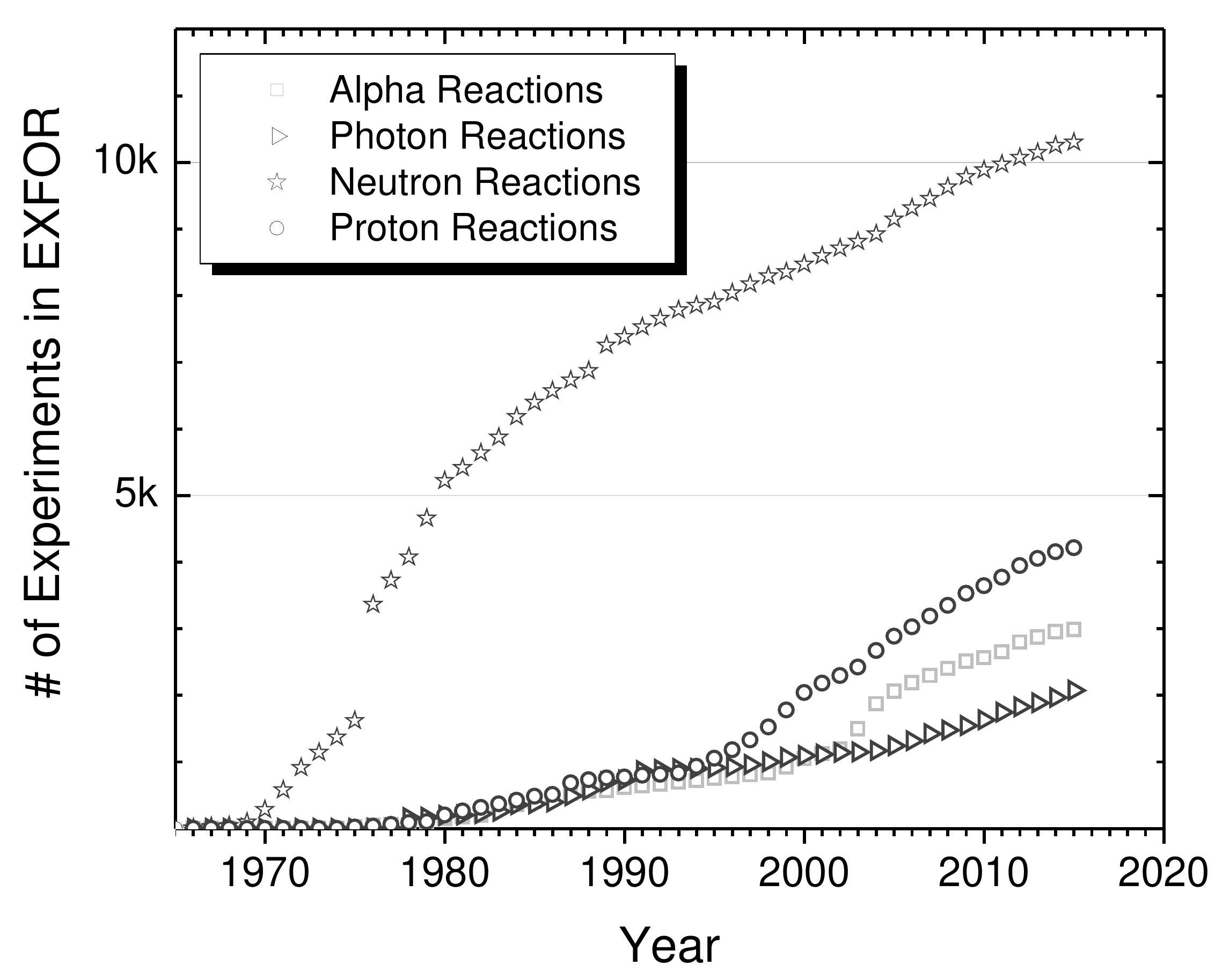}
\caption{Content evolution of the EXFOR database. Neutron reactions have been compiled systematically since the discovery of the neutron, while charged particle and photon reactions have 
been covered less extensively. %Initially the database scope was limited to neutron-induced reaction cross sections, later scope expansion included charge particle and photon reactions.
}
\label{fig:x4}
\end{figure}

Historically, EXFOR was created in support of nuclear energy research and development activities. 
The initial database scope was limited to neutron-induced reactions in order to provide support for ENDF evaluations \cite{Chadwick11}. 
Later, the scope was extended to charged-particle and photon-induced reactions. Presently, neutron-, proton-, alpha-, and photon-induced reactions 
constitute 47.9, 19.8, 7.36 and 6.2 $\%$ of database content, respectively. The historic evolution of EXFOR content is shown in Fig. \ref{fig:x4}.  

\begin{figure}[]
 \centering
 \includegraphics[width=0.7\textwidth]{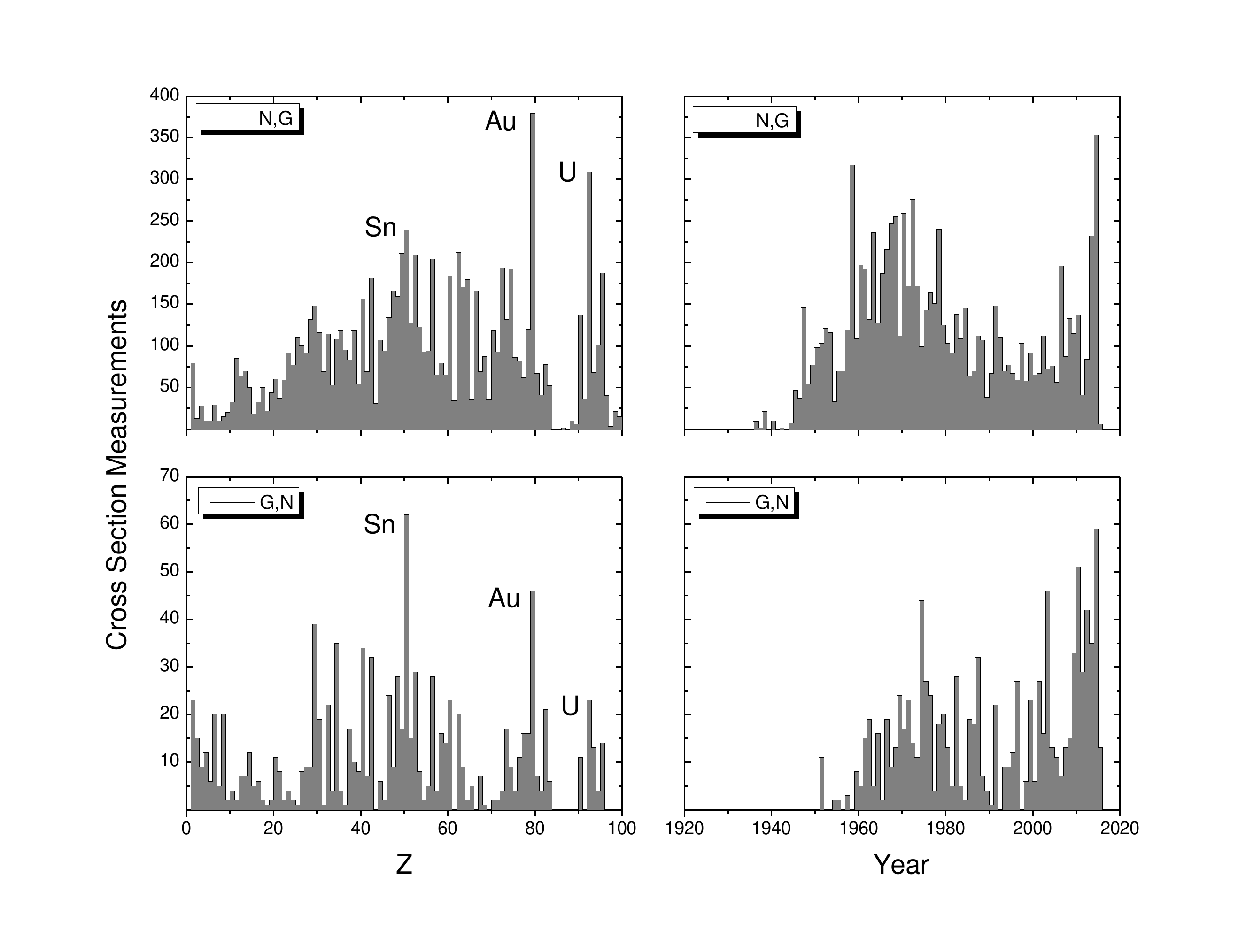}
\caption{EXFOR compilations of astrophysically-important neutron capture and photo-neutron production reactions vs. reaction target and publication year of experimental results.}
\label{fig:neutron}
\end{figure}

The EXFOR library contains a wealth of nuclear reaction data relevant to nucleosynthesis modeling because its present scope 
is broad and compilations are systematically performed in the last 50 years. To illustrate this point I will consider EXFOR data 
on the neutron capture and production reactions that play an essential role in {\it s}, {\it r} and {\it p} processes, respectively. 
The analysis of data shown in Fig. \ref{fig:neutron}  indicates that the most frequently-used neutron capture targets are gold, uranium and tin, 
while neutron-rich tin, gold and uranium are materials of choice for photo-neutron reaction studies. Further examination of data history provides 
indication of two trends: large numbers of nuclear reaction measurements in the 1970's \cite{Pritychenko15} 
and a recent increase in (n,$\gamma$) and ($\gamma$,n) reactions. 
The last trend shows clear proof of nucleosynthesis research impact on the nuclear reaction field in the last 15 years.
Direct results on Maxwellian-averaged cross sections and astrophysical reaction rates are also compiled in EXFOR.

The EXFOR project has a very advanced Web interface ({\it http://www.nndc.bnl.gov/exfor}) \cite{Zerkin05} that provides direct access to data and evaluation tools, 
such as an inverse reaction calculator which is based on the nuclear reaction reciprocity theorem. It allows extracting a reaction cross section if an inverse reaction is known. 
For 1 + 2 $\rightarrow$ 3 + 4 and 3 + 4 $\rightarrow$ 1 + 2 processes the cross section ratio is 
\begin{equation}
\label{myeq.r6}
\frac{\sigma_{34}}{\sigma_{12}} = \frac{m_3 m_4 E_{34}(2 J_3 + 1) (2 J_4 + 1)(1 + \delta_{12})}{m_1 m_2 E_{12}(2 J_1 + 1) (2 J_1 + 1)(1 + \delta_{34})},
\end{equation}
where $E_{12}$,  $E_{34}$ are kinetic energies in the c.m. system, J is angular momentum, and $\delta_{12}$=$\delta_{34}$=0. 
The EXFOR project is a primary source of experimental nuclear reaction data for many potential applications, 
such as neutron cross section evaluations. Next, I will discuss a relationship between EXFOR and ENDF databases.

\subsubsection{ENDF}

\begin{figure}[]
 \centering
 \includegraphics[width=0.7\textwidth]{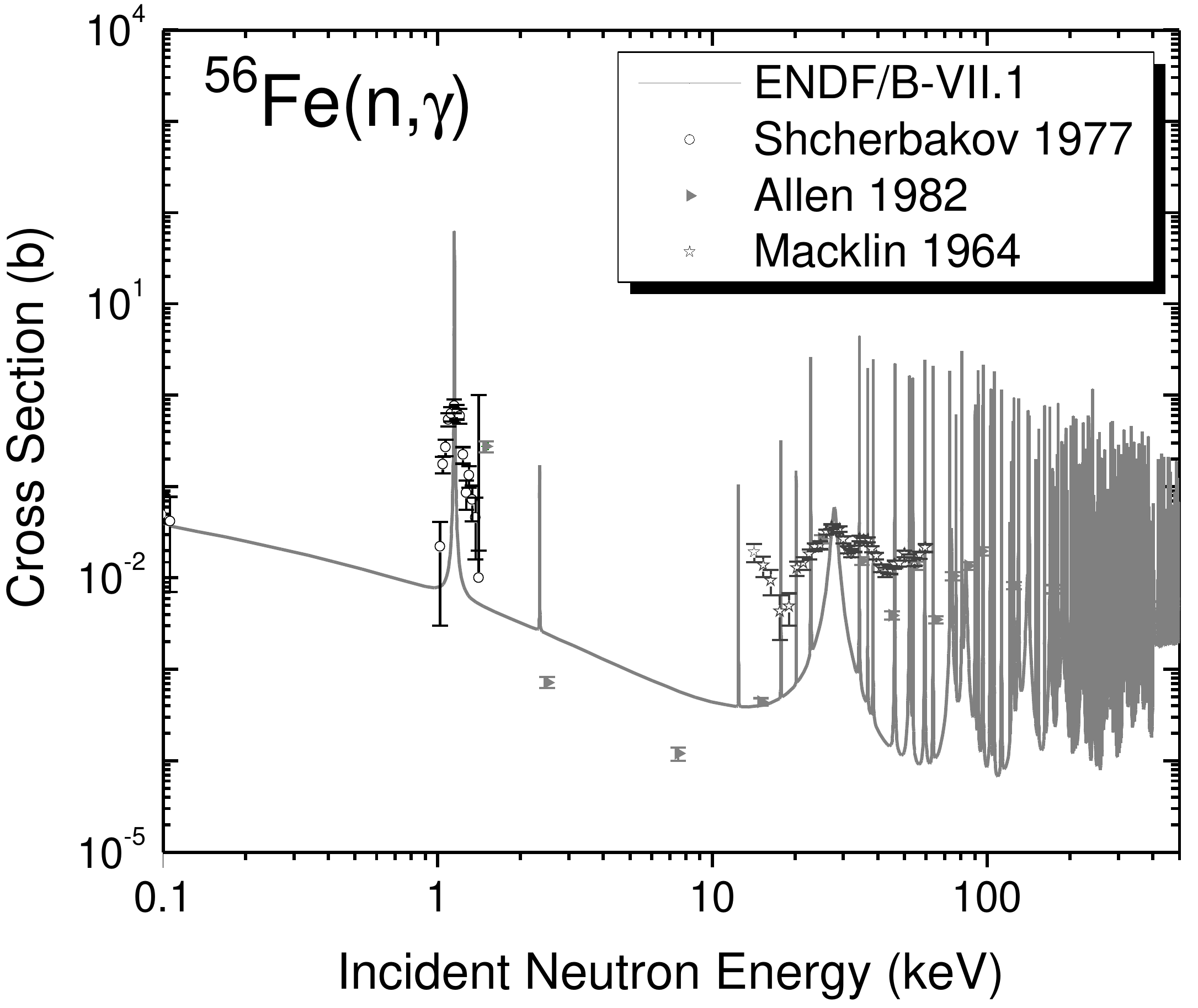}
\caption{ENDF/B-VII.1 library \cite{Chadwick11} evaluated and EXFOR library \cite{Otuka14} experimental    $^{56}$Fe(n,$\gamma$) cross sections for astrophysical range of energies.}
\label{fig:56Fe}
\end{figure}

 ENDF is a core nuclear reaction database containing evaluated (recommended) cross sections, 
spectra, angular distributions, and fission product yields, thermal neutron scattering, photo-atomic and other data, 
with an emphasis on neutron-induced reactions. ENDF library evaluations are based on theoretical calculations adjusted to experimental data  
with an exception of neutron resonance region, where priority is given to experimental data. 
ENDF evaluations consist of complete neutron reaction data sets for particular isotope or element (ENDF material). They cover all neutron reaction channels 
within the 10$^{-5}$ eV - 20 MeV energy range. All these data sets are stored in the internationally-adopted ENDF-6 format that is maintained by Cross Section Evaluation Working Group 
and publicly available at the National Nuclear Data Center (NNDC) website ({\it http://www.nndc.bnl.gov/endf}). Neutron cross sections for astrophysical 
range of energies, such as  the $^{56}$Fe(n,$\gamma$) reaction shown in Fig. \ref{fig:56Fe}, are available in the  ENDF/B-VII.1 \cite{Chadwick11} 
and EXFOR libraries. Cross section evaluations are of great importance for 
nuclear energy and science applications, and they are performed internationally for ENDF \cite{Chadwick11}, 
JEFF \cite{Ko11}, JENDL \cite{Shi11}, ROSFOND \cite{Za07} and CENDL \cite{Ge11}  libraries in the USA, Europe, Japan, Russian Federation, and China. 

Recent releases of  evaluated nuclear reaction libraries  have created a unique opportunity of applying 
these data for non-traditional applications, such as {\it s}-process nucleosynthesis \cite{BPritychenko10}.  The feasibility study of the evaluated neutron cross sections for {\it s}-process nucleosynthesis is presented below.

%\subsubsection{Calculation of Maxwellian-averaged Cross Sections and Uncertainties}
The Maxwellian-averaged cross section  can be expressed as   
\begin{equation}
\label{myeq.max3}
\langle \sigma^{Maxw}(kT)  \rangle = \frac{2}{\sqrt{\pi}} \frac{(m_2/(m_1 + m_2))^{2}}{(kT)^{2}}  \int_{0}^{\infty} \sigma(E^{L}_{n})E^{L}_{n} e^{- \frac{E^{L}_{n} m_2}{kT(m_1 + m_2)}} dE^{L}_{n},
\end{equation}
where {\it k} and {\it T} are the Boltzmann constant and temperature of the system, respectively,  and $E$ is an energy of 
relative motion of the neutron with respect to the target. Here,  $E^{L}_{n}$ is a neutron energy in the laboratory system 
and $m_{1}$ and $m_{2}$ are masses of a neutron and target nucleus, respectively.

The astrophysical reaction rate for network calculations is defined as
\begin{equation}
\label{rrate}
R(T_9) = N_{A}\langle\sigma v\rangle = 10^{-24}  \sqrt(2 kT/\mu)  N_{A}  \sigma^{Maxw}(kT),
\end{equation}
where $T_9$ is temperature expressed in billions of Kelvin, $N_A$ is an Avogadro number. $T_9$ is related to the   $kT$ in  MeV units as follows
\begin{equation}
\label{t9}
11.6045 \times T_9 = kT
\end {equation}

It is commonly known that  for the equilibrium {\it s}-process-only nuclei product  of $\langle \sigma^{Maxw}_{\gamma} (kT) \rangle$ and solar-system abundances ($N_{(A)}$)  is preserved \cite{Rolfs88}
\begin{equation} 
\label{myeq.s0} 
\sigma_{A}N_{(A)}= \sigma_{A-1}N_{(A-1)} = constant
\end{equation}

The stellar equilibrium conditions provide an important test for  {\it s}-process nucleosynthesis in mass regions between neutron magic 
numbers N=50, 82, 126 \cite{Arlandini99}. To investigate this phenomenon,  I will consider  ENDF/B-VII.1  library cross sections. 
These data were never fitted for nuclear  astrophysics models and  are essentially model-independent.

\begin{figure}[htb]	% h-here, t-top, b-bottom
\centering
\includegraphics[width=0.7\textwidth]{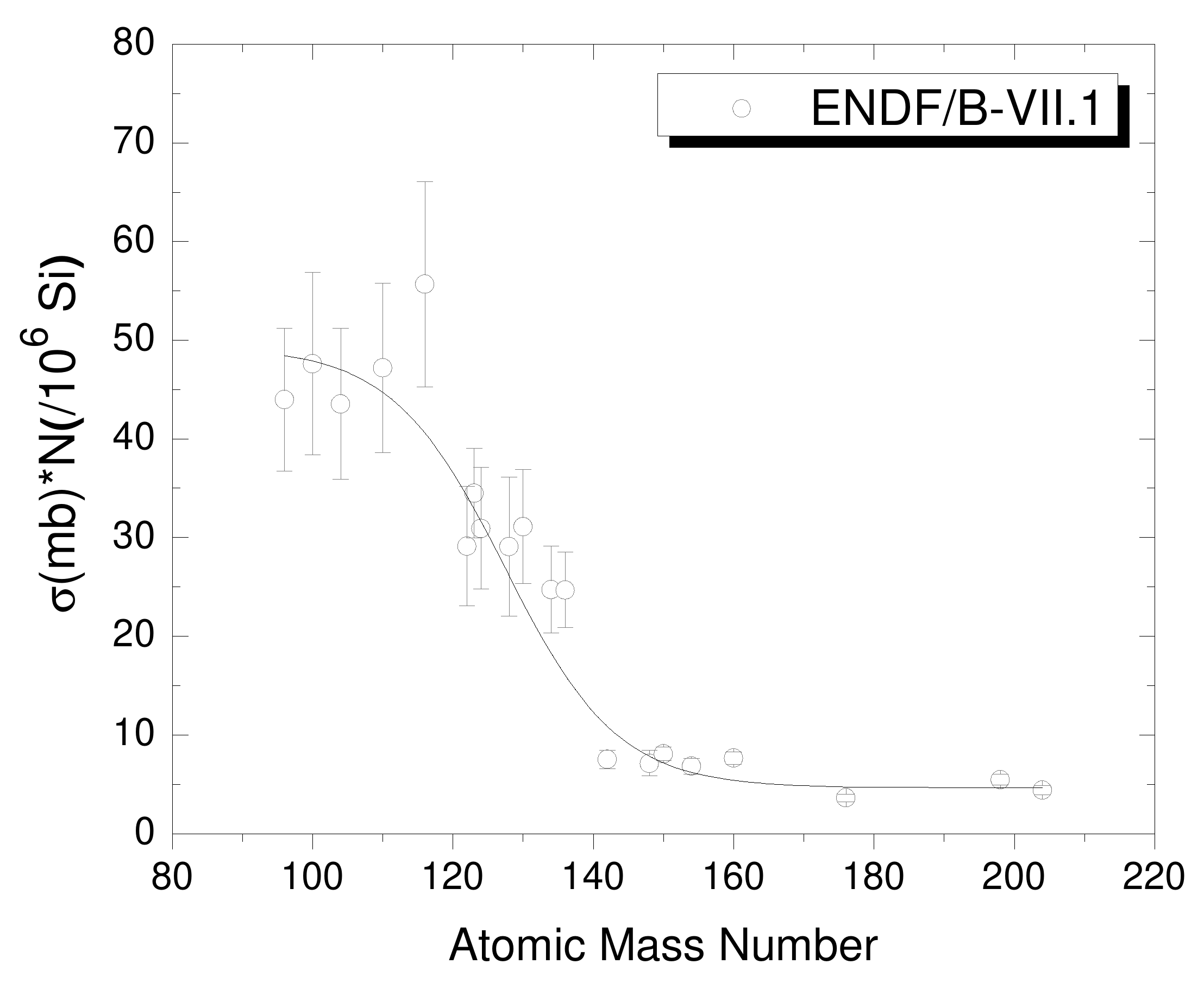} %	** if .eps don't need extension
\caption[ENDF/B-VII.1 library product of neutron-capture cross section (at 30 keV in mb) times solar system abundances (relative to Si = 10$^6$) as a function of atomic mass for nuclei produced only in the {\it s}-process.]{ENDF/B-VII.1 library product of neutron-capture cross section (at 30 keV in mb) times solar system abundances (relative to Si = 10$^6$) as a function of atomic mass for nuclei produced only in the {\it s}-process.}
\label{fig:2Plate}
\end{figure}

The product values of the ENDF/B-VII.1 $\langle \sigma^{Maxw}_{\gamma} (30 keV) \rangle$ times solar abundances \cite{Anders89} are plotted  in Fig.  \ref{fig:2Plate}.  
They reveal that the ENDF/B-VII.1 library data closely replicate a two-plateau plot \cite{Rolfs88} and provide a powerful testimony for stellar nucleosynthesis.

The predictive power of stellar nucleosynthesis calculations depends heavily on  neutron cross section values and their covariance matrices. 
To understand the unique isotopic signatures from the presolar grains, $\sim$1 $\%$ cross section uncertainties are necessary \cite{FKaeppeler11}. 
Unfortunately, present experimental and theoretical errors are often much higher.

\subsection{Nuclear Structure and Decay Data}

\begin{figure}[htb]	% h-here, t-top, b-bottom
\centering
\includegraphics[width=0.7\textwidth]{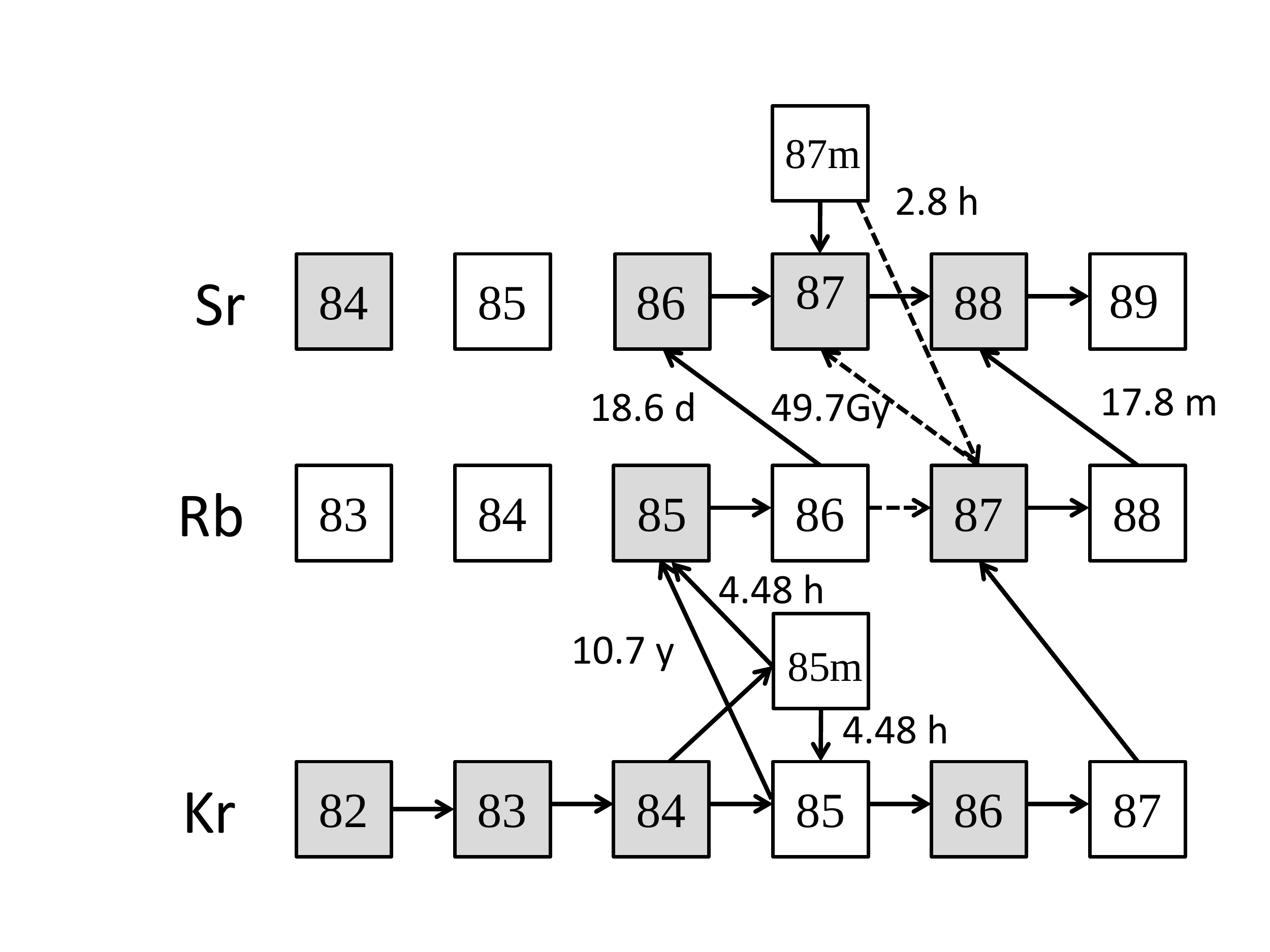} %	** if .eps don't need extension
\caption{Fragment of the {\it s}-process path: The  Kr-Rb-Sr  region  of  the  chart  of  the  nuclides. Stable  isotopes  are  in  shaded  boxes,  unstable  isotopes  in  open boxes.   
The major and minor channels for neutron capture and $\beta$-decays are indicated by arrows (solid and dashed).}
\label{fig:SPath}
\end{figure}

Analysis of the {\it s}-process path in the Kr-Rb-Sr  region indicates that cross section data alone cannot explain the present-day variety of elements and isotopes. 
The data in Fig. \ref{fig:SPath}  clearly show that decay information is absolutely essential in stellar nucleosynthesis modeling. 

\subsubsection{ENSDF}

 The best source of decay data is the ENSDF database \cite{Burrows90,ensdf16}, and the  Kr-Rb-Sr data sets are shown in Table \ref{tbl1}.   
 ENSDF contains evaluated structure and decay data sets for over 3000 nuclides as shown in Fig. \ref{fig:chart}, and it  is predominantly a USNDP product 
with some contributions from Europe and Asia.

\begin{table*}
\begin{center}
\caption{ENSDF decay data sets for Kr-Rb-Sr  region of nuclei. These data are publicly available from ENSDF Web interface ({\it http://www.nndc.bnl.gov/ensdf}) \cite{ensdf16}.\label{tbl1}}

\begin{tabular}{c|c|c|c}
\hline\hline
Isotope	& Decay Mode      & 	Branching Ratio & 	Half life \\
        &                 &     $\%$         &                 \\
\hline
$^{85m}$Kr  &	IT	  & 	21.2(6)  & 	4.480(8) h  \\
$^{85m}$Kr  &	$\beta$-  &	78.8(6)	 &      4.480(8) h  \\
$^{85}$Kr   &	$\beta$-  &    	100	 &     10.739(14) y \\
$^{87}$Kr   &	$\beta$-  &	100	 &     76.3(5) m \\
$^{87m}$Sr  & 	EC	  &     0.30(8)	 &     2.815(12) h \\
$^{86}$Rb   &  	$\beta$-  &   99.9948(5) &    18.642(18) d \\
$^{87}$Rb   &	$\beta$-  &	100	 &    4.97(3)$\times$10$^{10}$ y \\
$^{88}$Rb   &	$\beta$-  &	100	 &    17.773(18) m  \\
\hline\hline
\end{tabular}
\end{center}
\end{table*}

The ENSDF library is essentially based on evaluation of experimental data,  and all  data for a given mass chain or nuclide are re-evaluated on a 10-12 year basis. 
It uses the mathematical procedures and codes to deduce the evaluated or recommended values.  
ENSDF decay data sets contain the following information \cite{Burrows90}:
\begin{itemize}
\item Parent level energy, J$^{\pi}$, T$_{1/2}$ and Q-value for the parent ground-state to daughter ground-state decay.
\item Branching ratio for the decay mode and normalization factors for converting the relative emission probabilities for $\gamma$-ray and $\beta$- or electron capture to absolute emission probabilities.
\item $\gamma$-ray and E0 transition energies, emission probabilities, multipolarities and mixing ratios and the total and partial internal conversion coefficients.
\item $\alpha$-, $\beta$- and electron capture transition energies.
\end{itemize}

These ENSDF decay data play an important role at stellar nucleosynthesis  branch points where $\beta$-decay and neutron capture rates are 
\begin{equation}
\label{myeq.br}
\lambda_{n} \sim \lambda_{\beta}
\end{equation}
At these points, the stellar thermal environment may complicate the situation even further by changing  $\beta$-decay rates and affecting the process path. Branching is particularly important 
for unstable isotopes such as  $^{134}$Cs. This isotope can either decay to $^{134}$Ba, if neutron flux is low or capture neutron and produce $^{135}$Cs. 
The  $^{134}$Cs $\beta$-decay lifetime may vary from $\sim$1 year to 30 days over the temperature range of (100-300)$\times$10$^6$ K. These variations are due to 
differences between decay rates of neutral and highly-ionized atoms \cite{Takahashi87}. In many cases where experimental data are not available, nuclear theory  
may provide complete data sets for stellar nucleosynthesis calculations \cite{Moller95,Moller97}.

\section{Data Evaluation Codes and Procedures}

Over the last 65 years, the USNDP has accumulated a wealth of experimental knowledge and a diverse range of data analysis and evaluation techniques. The nuclear reaction 
data evaluations often use EMPIRE, TALYS and Atlas collection of nuclear reaction model codes \cite{Herman07,Ko08,Mu06} 
that are frequently calibrated to fit experimental data \cite{Otuka14}. The model codes are essential for neutron cross section calculations of short-lived radioactive nuclei where experimental data are not available.

The structure and decay data evaluations are often based on direct evaluation of experimental results.  ENSDF data evaluation philosophy includes collection of the best available data from each type of experiment and nuclide in a concise and  well-documented manner \cite{Burrows90}. The emphasis in nuclear data evaluation is on conservative analysis of the structure and decay properties obtained from experimental evidence, 
well-founded systematics, or theory. The propagation of uncertainties is often problematic due to the large amount of data, the correlations between uncertainties, 
and  insufficient information provided by authors. These problems are further complicated by the need for correct propagation of large uncertainties, 
asymmetric uncertainties, limits, and ranges. A recent NNDC review  of data uncertainties \cite{Brown15} provides a complete description of this topic.  

Over the years many of these problems have been addressed, covariance formalism has been developed for ENDF libraries \cite{Chadwick11},  and new computer codes were introduced. 
The data analysis and evaluation codes are pure mathematical procedures that can be applied to diverse data samples. In this chapter, I will discuss a nontraditional application of 
nuclear data evaluation codes for Hubble constant experimental data sets evaluation. 

\subsection{Hubble Constant Experimental Data Sets}

\begin{figure}[htb]	% h-here, t-top, b-bottom
\centering
\includegraphics[width=0.7\textwidth]{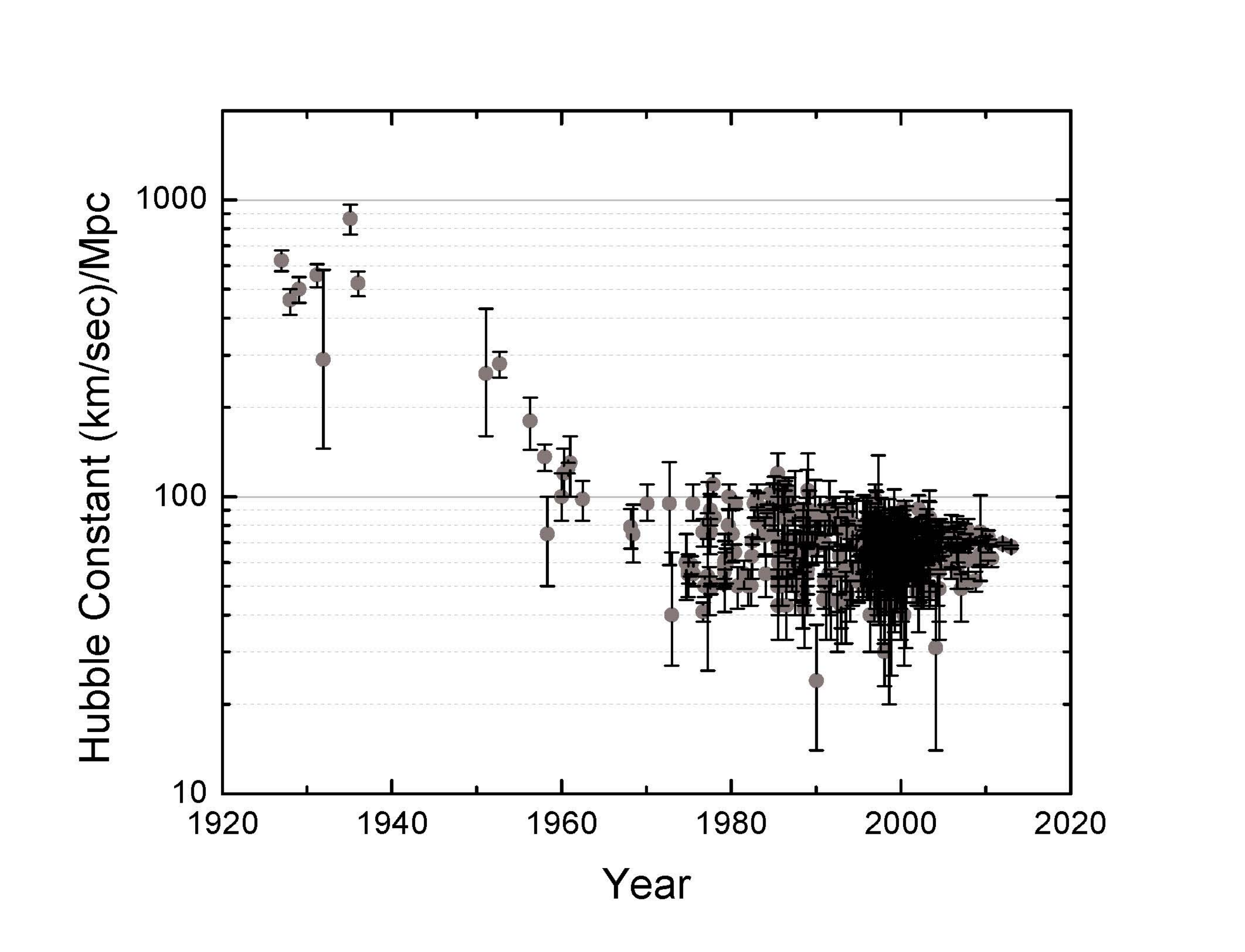} %	** if .eps don't need extension
\caption{Historical evolution of Hubble constant measurements.}
\label{fig:Hubble}
\end{figure}

The Hubble constant and its present-day numerical value are very important for modern astrophysics and 
cosmology \cite{Boyd08,Dolgov88}. For many years, researchers have been improving accuracy of the constant \cite{livio13}. 
Fig. \ref{fig:Hubble} shows the time evolution of Hubble research in the last 100 years, and the original effort can be traced to as early as 1916 \cite{huchra15}. 
The large number of measurements creates a certain degree of confusion about Hubble constant numerical value, and scientists often 
rely on recently-published results \cite{Olive14}. The precision of Hubble constant measurements has improved enormously over the years; 
however, it is not always prudent to reject older results in favor of the latest findings. 
Consequently, it makes perfect sense to analyze all available results, evaluate the data, and extract the recommended value.

\subsubsection{Hubble Constant Data Evaluation}

The volume of Hubble constant measurements far exceeds other experimental quantities in nuclear physics \cite{BPritychenko15} and supplies a large input sample for USNDP codes. 
Previously, similar situations have been encountered in nuclear and particle physics and resolved with data evaluations. 
Nuclear data evaluations and their policies are well described in literature \cite{pritychenko16,ensdf16}. 
Frequently, the evaluations are completely based on or adjusted to available experimental data. 
These adjustments and specialized mathematical statistics techniques can be applied for nuclear, particle, or any other data sets. 
In this work, I would follow standard nuclear data evaluation procedures to deduce the recommended value. Current evaluation input 
data are mostly based on the NASA/HST Key Project on the Extragalactic Distance Scale compilation \cite{huchra15} and recent results. 
A visual inspection of historical Hubble Constant measurements, as shown in Fig. \ref{fig:Hubble}, is instrumental in the data analysis. It suggests 
that one may safely reject all measurements prior to 1970. It is common knowledge that Hubble constant measurements heavily rely on 
the accuracy of astronomic distance determination. 

Older results, such as those reported by A. Sandage and G. de Vancouleurs, suffered from inaccurate measurements \cite{livio13}. 
Therefore, the rejection of all results prior to 1990 could provide a complimentary benchmark value of the Hubble constant. 
In the present data analysis, the experimental data have been separated into two groups, with 1970 and 1990 time cuts, 
and further reduced using the following policies:
\begin{itemize}
\item Rejection of repeated results (multiple publication of the same result).
\item Rejection of model-dependent results (i.e. Cosmic Microwave Background (CMB) fits).
\item Rejection of potential outliers using Chauvenet's criterion \cite{birch14}.
\end{itemize}
Common data evaluation practices indicate that recommended value should be based on a large statistical sample that includes different 
types of measurements. The 1970 and 1990 redacted data sets of $\sim$334 and $\sim$266 data points, respectively, provide such samples. 
These large samples create the possibility of deducing Hubble constant value for each method of observation besides the combined value 
that is based on all measurements. The current data collections were further subdivided using a NASA/HST Key Project on the 
Extragalactic Distance Scale classification of experimental methods \cite{huchra15}:
\begin{itemize}
\item S = Type Ia supernovae (SNIa).
\item 2 = Type II supernovae (SNII).
\item L = Lens.
\item r = Red Giants.
\item B = Baryonic Tully-Fisher.
\item R = Inverse Tully-Fisher (ITF, RTF).
\item H = Infrared Tully-Fisher (or IRTF).
\item F = Fluctuations.
\item A = Global Summary.
\item Z = Sunyaev-Zeldovich.
\item T = Tully-Fisher.
\item O = Other.
\end{itemize}

These experimental data sets have been processed with the latest version of the visual averaging library \cite{birch14}. 
The library program includes limitation of relative statistical weight (LWM), normalized residual (NRM),
Rajeval technique (RT), and the Expected Value (EVM) statistical methods to calculate averages of experimental 
data with uncertainties. The experimental data sets were processed, and evaluated values with reduced $\chi^2$$<$2 were typically 
accepted as reasonable data fits. The current evaluation incorporates statistical methods based on the inverse squared value 
of the quoted uncertainties, a procedure that is consistent with the general methodology used in treatment of data 
for the ENSDF database \cite{ensdf16} and Particle Data Group \cite{Olive14}.

\subsubsection{Hubble Results and Discussion}

\begin{figure}[htb]	% h-here, t-top, b-bottom
\centering
\includegraphics[width=0.7\textwidth]{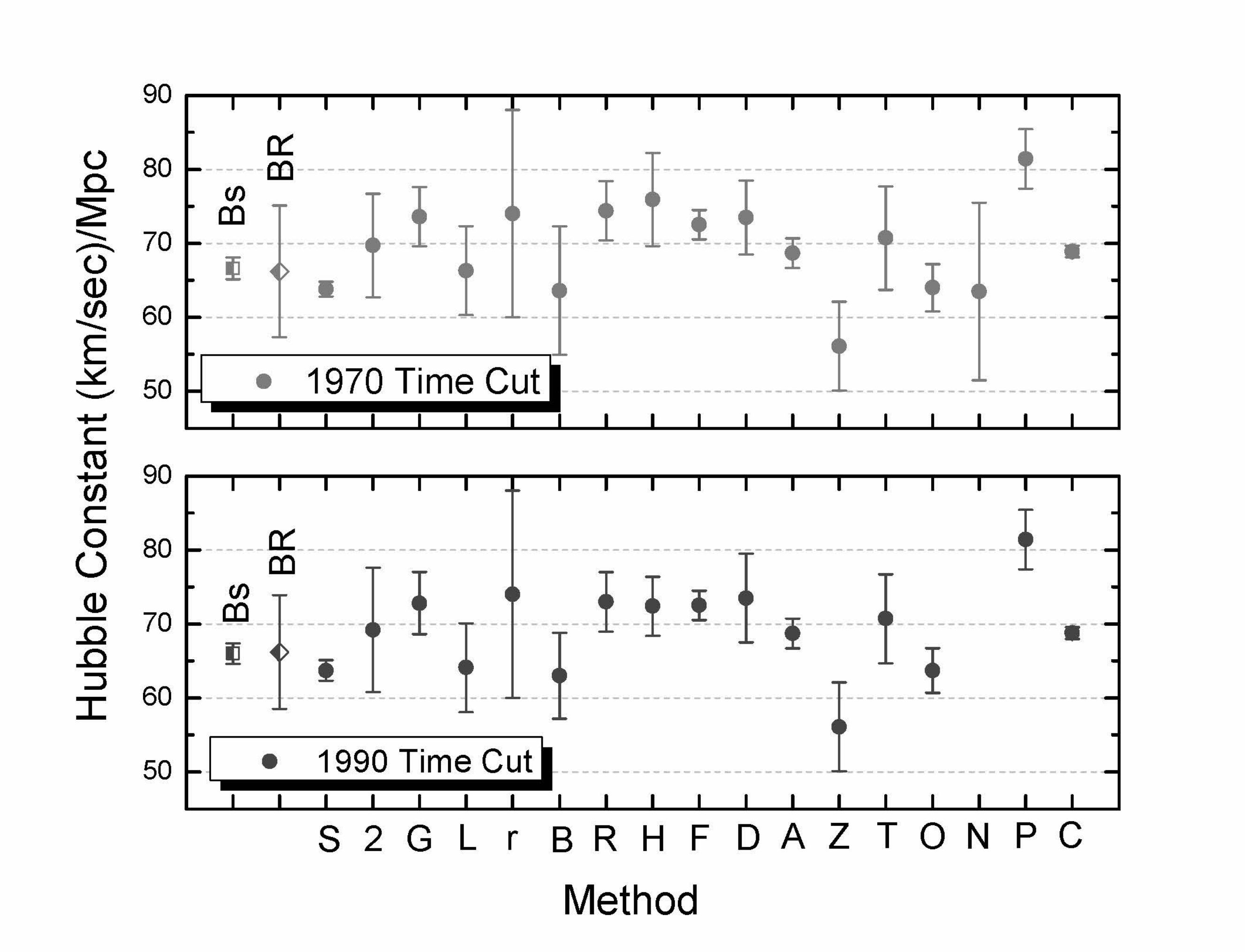} %	** if .eps don't need extension
\caption{Evaluated values of Hubble constant using 1970 and 1990 time cuts for experimental data. These plots include the evaluated or recommended values for combined observations, using the Best Representation (BR) and Bootstrap (Bs) data analysis techniques. Globular cluster luminosity function (G), Dn-Sig/Fund
Plane (D), Novae (N), Planetary nebula luminosity function (P) results are also included while CMB Fit(C) value is shown for comparative purposes.}
\label{fig:HubbleCuts}
\end{figure}

Two sets of recommended values are displayed in Fig. \ref{fig:HubbleCuts}, and the combined numerical values are shown in Table \ref{tbl2}. 
The Hubble constant combined central values extracted by means of different mathematical techniques are in good agreement, 
while uncertainties need further discussion.

\begin{table*}
\begin{center}
\caption{Results of the Hubble Constant evaluation for all observations using 1970 and 1990 time cuts.\label{tbl2}}

\begin{tabular}{c|c|c}
\hline\hline
Method &	Time Cut: 1970	 & Time Cut: 1990 \\
              & (km/sec)/Mpc & (km/sec)/Mpc \\
\hline
Unweighted Average &	67.56(73)	& 65.92(65) \\
Weighted Average & 63.68(58)	 & 63.87(51) \\
LWM &	67.3(93)	& 66.08(64) \\
Normalized Residual & 64.50(50) &	64.33(48) \\
Rajeval Technique &	65.57(45) & 	65.07(46) \\
Best Representation (BR) &	66.2(89)	 &  66.2(77) \\
Bootstrap (BS) & 	66.6(15)	& 66.0(14) \\
Mandel-Paule &	66.7(92)	& 65.3(62)\\
\hline\hline
\end{tabular}
\end{center}
\end{table*}

Finally, two different time cuts of 1970 and 1990 for Hubble's data have yielded two recommended values of 66.2(89) and 66.2(77) (km/sec)/Mpc, respectively. 
The agreement between these values partially reflects the fact that the majority of the Hubble's constant measurements have been performed in the last 25 years, 
and a small number of potential outliers has been rejected. More accurate recent observations imply a preference for the 1990 time cut 
value of 66.2(77) (km/sec)/Mpc. The last result is consistent with the Hubble Space Telescope and Wide Field Camera 3 \cite{Riess11} and 
the model-dependent Planck's Mission and WMAP values \cite{ade13,bennett13}. 
Inclusion of the globular cluster luminosity function, Dn-Sig/Fund Plane, Novae, and planetary nebula luminosity function data would slightly 
change 1970 and 1990 recommended values to 66.9(90) and 66.9(78), respectively. These less precise measurements do not affect 
the recommended values severely because of a Chauvenet's criterion analysis.

In recent years, an effort has been made to calculate the Hubble constant median statistics \cite{chen11}. The median statistics approach 
differs substantially from  nuclear or particle data evaluation procedures. At the same time, it provides a complementary value of 
68$\pm$5.5 (or $\pm$1) (km/sec)/Mpc, where the errors are the 95$\%$ statistical and systematic (or statistical) errors, 
that can be compared with the present results.

The knowledge of Hubble constant value has multiple implications in science. As an example, a rough estimate of the age of the Universe can be 
deduced using the standard methodic \cite{wiki15}. The adopted value of 66.2(77) (km/sec)/Mpc implies (14.78$\pm$1.72)$\times$10$^{9}$ years estimated 
value for the age of Universe. The last result is consistent with the recently published value of (13.798$\pm$0.037)$\times$10$^{9}$ years \cite{ade13}.

The analysis of Hubble constant measurements has been performed using standard USNDP codes and procedures. An evaluated data set of most probable 
values of Hubble constant has been deduced and shown in  Table \ref{tbl2}. These values are consistent with other available results. 
An accurate constant value is instrumental for many potential applications. The recommended value of the constant is completely based 
on experimental measurements, and further, more precise observations, would lead to a more accurate determination of it.

\subsection{Further Analysis of Astrophysical Data Sets}

In the near future,  data analysis and evaluation codes could be applied to experimental nuclear reaction data in order to produce 
data sets of cross section values and uncertainties  complementary to astrophysical libraries \cite{Angulo99,Cyburt10,Dillmann06}.

\section{Conclusion}

A brief review of astrophysical modeling data needs has been presented with an emphasis on nucleosynthesis modeling. 
The scope of the traditional nuclear astrophysics activities has been extended to the USNDP nuclear databases, codes, and methods.  
The EXFOR and ENDF libraries are extensively investigated in the context of neutron cross sections for  {\it s}-process nucleosynthesis.  
Further analysis shows that decay rates are essential in stellar nucleosynthesis calculations, 
and these rates have been evaluated for ENSDF database using nuclear structure codes. 

The nuclear data evaluation codes and procedures have been applied to Hubble constant measurements. This approach results in  
the most probable or recommended Hubble constant value of 66.2(77) (km/sec)/Mpc and   implies (14.78$\pm$1.72) $\times$ 10$^{9}$ years 
as a rough estimate for the age of the Universe. 

The current analysis of astrophysical nuclear data needs indicates a strong potential of USNDP databases for  astrophysical applications. Further work 
may produce results in complementary to astrophysics libraries data sets. These data could play a crucial role in stellar nucleosynthesis modeling and other 
potential applications. 

\bigskip

\noindent{\bf Acknowledgements:}
%\begin{acknowledgements}
The author is indebted to Dr. M. Herman (BNL) for support of this project and grateful to Dr. V. Unferth (Viterbo University) for help with the manuscript. 
This work was funded by the Office of Nuclear Physics, Office of Science of the U.S. Department of Energy, under Contract No. DE-AC02-98CH10886 with Brookhaven Science Associates, LLC.
%\end{acknowledgements}

%\bibliographystyle{spbasic}
\bibliographystyle{spphys}       % APS-like style for physics
%\bibliography{FBpapers}

\label{lastpage-01}

\end{document}